\documentclass{elsart}
\usepackage{epsf,amssymb,natbib}
\usepackage{times}      
\journal{Planetary and Space Science -- special issue}

\newcommand{\DDS}{\renewcommand{\baselinestretch}{1.5}\tiny \normalsize}

\newcommand{\SSM}{\renewcommand{\baselinestretch}{0.8}\tiny \normalsize}

\newcommand{\AU}{{\footnotesize AU}}

\newcommand{\kms}{\mbox{km\,s$^{-1}$}}

\newcommand{\rj}{\mbox{$R_{\rm J}$}}
    \def\independenT#1#2{\mathrel{\setbox0\hbox{$#1#2$}%
    \copy0\kern-\wd0\mkern4mu\box0}} 
    
\newcommand{\Lower}[1]{\smash{\lower 1.5ex \hbox{#1}}}

\newlength{\captsize}   \let\captsize=\footnotesize
\newlength{\captwidth}  
\newlength{\beforetableskip} 
\newcommand{\capt}[1]{\begin{minipage}{\captwidth}
\let\normalsize=\captsize
\caption[#1]{#1}
\end{minipage}\\ \vspace{\beforetableskip}}

\makeatletter

\begin{document}

\begin{frontmatter}

\title{Characterization of Jovian Plasma-Embedded Dust Particles}
\author{Amara L. Graps}
\address{INAF Istituto di Fisica dello Spazio Interplanetario, CNR-ARTOV,
Via del Fosso del Cavaliere 100, 00133 Rome, Italy. Fax: +39-06-4993-4383}
\ead{ amara.graps@ifsi-roma.inaf.it}
\date{}
%


\begin{abstract}

As the data from space missions and laboratories improve, a research
domain combining plasmas and charged dust is gaining in prominence.
Our solar system provides many natural laboratories such as planetary
rings, comet comae and tails, ejecta clouds around moons and
asteroids, and Earth's noctilucent clouds for which to closely study
plasma-embedded cosmic dust. One natural laboratory to study
electromagnetically-controlled cosmic dust has been provided  by the
Jovian dust streams and the data from the instruments which were on
board the Galileo spacecraft. Given the prodigious quantity of dust
poured into the Jovian magnetosphere by Io and its volcanoes resulting
in the dust streams, the possibility of dusty plasma conditions exist.
This paper characterizes the main parameters for those interested in
studying dust embedded in a plasma with a focus on the Jupiter
environment.  I show how to distinguish between dust-in-plasma and
dusty-plasma and how the Havnes parameter $P$ can be used to support
or negate the possibility of collective behavior of the dusty plasma.
The result of applying these tools to the Jovian dust streams reveals
mostly dust-in-plasma behavior. In the orbits displaying the highest
dust stream fluxes, portions of orbits E4, G7, G8, C21 satisfy the
minimum requirements for a dusty plasma. However, the $P$ parameter
demonstrates that these mild dusty plasma conditions do not lead to
collective behavior of the dust stream particles.
\end{abstract}

\begin{keyword}
charged dust, Jovian dust streams, dusty plasma, Jovian magnetosphere, Havnes parameter,
Debye length, Galileo mission
\end{keyword}

\end{frontmatter}



\section{Introduction}

As the data from space missions and laboratories improve, a research
domain combining plasmas and charged dust is gaining in prominence.
Our universe is comprised of 99\% plasma, and cosmic dust is
ubiquitous, yet a discussion of the interplay between the two is
often missing from the astrophysics textbooks. Our solar system
provides many natural laboratories such as planetary rings, comet comae and tails,
ejecta clouds around moons and asteroids, and Earth's noctilucent clouds 
for which to closely study plasma-embedded cosmic dust. 
After the Voyager spacecraft(s) left the giant planets, and the spacecraft flotilla
studying comet Halley went their own way too, one space laboratory
for studying dust-embedded-in-plasma appeared upon the arrival of
the Galileo spacecraft into the Jupiter magnetosphere. The
spacecraft carried an impact ionization dust instrument (Dust
Detector Subsystem or DDS) to sample in-situ the dust particles, and,
combined with the plasma instruments, an eight-year (1996-2003)
mission, and a strong dust source of the moon Io, from which material 
escapes at an approximate rate of 1~ton~sec$^{-1}$
\citep{Spens:96}, the potential to study dusty plasmas was enormous.

This paper characterizes the main parameters for those interested in
studying dust embedded in a plasma with a focus on the Jupiter
environment. I start with mathematical descriptions to aid one in
distinguishing between different dust-in-plasma regimes, I apply
these terms for the Jovian dust streams in order to identify in
which regions of the Jovian magnetosphere I might see collective
dust particle behavior, and then I apply these parameters. This
Jovian dust data comprises one population of several different
Jovian populations. The Jovian dust streams are high-rate bursts of
fast (at least 250 km-sec$^{-1}$) submicron-sized particles from Jupiter's
moon Io, in particular, dust from Io's volcanoes
\citep{Graps:2000a}. The plasma parameters used in the subsequent
tables were mostly derived from a model of the Jovian plasma.

To describe the dust particles' physical behavior, one must identify 
the involved forces. The intermediate force range between
gravitationally-dominated dust stream particles and 
electrodynamically-dominated ions and electrons is the regime where 
dust particles lie. Charged cosmic dust dynamics brings both forces
into play: particles typically micron-sized and larger are dominated
by gravity, while submicron-sized particles are progressively
dominated by electromagnetism (the Lorentz force), as their sizes
decrease. In order to quantify possible collective effects, I must
characterize {\it both} the Jovian plasma and the dust streams
particles, deriving their interparticle distances, energies, and
dynamics.

In dust-plus-plasma mixtures, there are three characteristic length
scales: 1)~the dust grain size $a$, the 2)~the plasma Debye length
$\lambda_D$, and 3) an average intergrain distance $d$, which is
equal to the dust number density $n_d$ by \mbox{$n_d d^3=1$}. 
In order to determine whether collective processes are
important, I use these length scales to compare plasma-plus-dust
regimes and for additional information, I calculate the Havnes parameter
$P$  \citep{Ver:2000,Goertz:89,Havnes:87}, which
is the ratio of dust charge density to the electron charge density,
given by $Z_dn_d/n_e$, where $n_e$
is the electron density and $Z_d$ is the dust particle charge number
$\left| {\rm Q} \right|/e$)
 for the same conditions.

The Debye screening length, $\lambda_D$ provides our first
indication whether collective processes might be important. The
Debye length is the distance over which the Coulomb field of
an arbitrary charge in the plasma is shielded. If the mean
separation between the dust particles in the plasma $d$ is smaller
than the Debye length, then neighboring charged dust particles are
not shielded and isolated from each other, and they begin to act
like a solid dielectric \cite[pg. 35]{Crav:97}.

Proceeding further to characterize collective processes. In the
dust-plus-plasma mixture, I can have two regimes depending on the
concentration of the dust grains:~\cite[Pg. 5-6]{Ver:2000}

 \begin{itemize}

 \item `\underline{Dust-in-Plasma}':  If $a \ll \lambda_D < d$, then
 I have a plasma containing isolated screened dust grains and can
 treat the dust from a particle dynamics point of view.

 \item `\underline{Dusty Plasma}': If $a \ll d < \lambda_D$, then
 collective effects of the charged dust can be relevant.

 \end{itemize}

\citet{Ver:2000}\ describes the complex charging of dust in a dusty plasma.
If the dust density in the plasma is increased, with an inverse
effect on the average distance $d$ between the grains, the
equilibrium charge on the particles decreases dramatically, and two
effects play a role in opposite directions 
\citep{Ver:2000,Goertz:89}:
grain capacitance increases, and each grain does not have to become
so negative (or positive), in order to equalize the ion (or electron)
currents to its surface.

The Jupiter's plasma discussed here, and in my dust stream dynamics
calculations is a fit by M. Hor\'anyi to the Voyager~1 and 2 cold plasma
measurements described in \citet{Bag:89}. Hor\'anyi assumed a
constant mixing ratio of 1:2 of singularly ionized oxygen to sulphur ions.
The plasma is quasi-neutral, i.e. $n_i$~=~$n_e$. I selected
particular locations: 6.2, 8, 10, 15, 20, 30~\rj\ (\rj~=~radial 
distance~71492~km from the center of Jupiter)
and present them in Table~\ref{plasmareps}\marginpar{Table~\ref{plasmareps}}. 
These particular locations, and temperatures ($T_e$ and $T_i$, respectively), electron 
density ($n_e$), and average ion mass numbers (avim in a.m.u.) will
be used again in subsequent tables.

In the following text, I provide equations (SI units) for the
individual particle (electrons, ions, dust particles) motions of the
dust-plus-plasma, for bulk motions of the plasma, and tables of the
values for the plasma model.

\section{Single Particle Motion}

In a constant magnetic vertical field $B_z$, the motion of an individual,
charged particle (electron, ion, tiny dust particle) is a helix of
constant pitch  around magnetic field lines. 
For this particle, the 
vertical velocity component $v_z$ is unaffected by 
$B$, while the  rotational velocity component: $v_\theta$ (=$\sqrt {v_x^2+v_y^2}$), 
is constant and gives the circular 
motion about the center of the orbit. Then:

$${ qv_\theta B ={{m_{(i,e,d)}v_{\theta}^2} \over {R_g}} \quad , \quad
R_g ={{m_{(i,e,d)} v_{\theta}} \over {qB}}\quad . \quad } \eqno(1) $$

where $q$ is the charge of the particle,  $v_\theta$ (=$\sqrt {v_x^2+v_y^2}$) is the rotational velocity component,
$m_{i,e,d}$ is its mass (where the subscript refers to either ions, electrons or dust
particles), and $R _g$ is its gyroscopic radius.
The gyroscopic radius can be associated with the temperature $T$
of the particle by using the thermal kinetic energy. Since
$v_z$ is unaffected by $B$, the rotational velocity is: 

$$v_{\theta}=\sqrt {{{2 k_BT} \over m_{(i,e,d)}}} \quad ,    \eqno(2) $$

\noindent \noindent where $k_B$ is the Boltzmann constant. Then the 
root-mean-squared gyroscopic radius in terms of temperature is: 

$$R_g={1 \over {qB}}\sqrt {2m_{(i,e,d)}k_BT}  \quad .  \quad  \eqno(3) $$

Note that the thermal energy of the particle becomes a part
of two constants of the motion: the total (vertical and rotational) 
kinetic energy ${1 \over 2}(m\vec v)$, 
and the magnetic moment: $({1/2}){({mv_\theta ^2}) / B}$, since 
$v_z$, $v_\theta$ and the mass of the particle is constant. 
The vertical velocity and hence its motion enters
via \mbox{$v_z=\sqrt{(2/m) (W - \Phi)}$}, where $W$
equals the total energy and $\Phi$ is the potential energy. 
An exploration of the
transfer of vertical kinetic energy in the context of charged
particles trapped in a drift tube is in \citet{Graps:1991}.

Another useful expression for the gyroradius is in terms of the gyrofrequency 
(also called cyclotron frequency or Larmor frequency):

$$R_g={ v_{\theta} \over \Omega_{(i,e,d)}} \quad , \quad \eqno(4) $$

where the magnitude of the
gyrofrequency $\Omega$ for ions, electrons and dust 
in SI units \cite[pg. 44]{Crav:97} is:

$$\Omega_{i,e,d} = { {q B} \over m_{(i,e,d)} }  \quad  {\rm (radians-s^{-1})} .  \quad \eqno(5) $$

\noindent For these purposes, I can consider a dipole for
Jupiter's magnetic field, $B = $ \mbox{$B_0 (\rj / r)^3$}, where $B_0$~=~4.28~G
is the field strength at the equator at Jupiter's surface~(\rj),
and $r$ = the radial distance from the center of Jupiter.

In Table~\ref{plasmasing1}, I list ion, electron and electron velocities, and dust sizes,
charges, velocities (average: $\langle v\rangle=v_\theta$), or speeds for the single particles. For the
ions and electrons, the velocities were calculated from the thermal kinetic energy (Eqn.~2),
given the masses ($m_i$, $m_e$) and the ion and electron temperatures ($T_i$, $T_e$) listed in
Table~\ref{plasmareps}\marginpar{Table~\ref{plasmasing1}}. These same parameters for 
the dust particles were gained by a different method.

The Galileo DDS did not measure the
dust particle charge directly, but measured, instead, an impact plasma that depended on 
the particle's velocity and mass, upon which laboratory
calibrations were applied. Therefore, for the dust  particles' sizes, charges, and velocities, I relied on 
the dust streams model in \citet{Graps:2001}\ to determine these properties indirectly. Investigators since 
1995 \citep{Gru:96b,Gru:96a,Gru:97,Hor:97,Graps:2001,Kru:2003c},
have best matched the fluxes or dynamics of the Jovian dust streams with a (seven free parameters)
model that includes  electron and ion collection currents as well as photoelectron emission 
and secondary electron emission currents. These currents were integrated to determine the
particle's charge simultaneously with the particle's dynamics. To generate average particle properties 
throughout the Jovian magnetosphere, I launched 287200 trajectories at Keplerian speeds from a ring just outside
Io's orbit at 6.2~\rj, and stopped the simulation after fifteen hours of dust particle
travel time. The particles 
that remained inside of the magnetosphere, I then considered to be representative for
many parameters of interest (with the exception of number density) of 
the general population of Jovian dust stream particles within 50~\rj.
For each dust particle trajectory,  I choose randomly between a minimum and maximum value 
for the free parameters:~1)~spherical particle  density and radius, 2)~radiation pressure 
scattering coefficient,  3)~photoelectric yield for the photoelectron emission current and 
the energy (Maxwellian)
distribution of photoelectrons released from grain when a photon
impacts onto the dust particle, 4)~maximum yield of secondary
electrons released by a high energy impacting electron or ion, 
and 5)~orbital phase, 6)~magnetic field orientation, and 7)~initial dust charge. 
Figure~1\marginpar{Figure~1} displays
the simulation output for the dust particles' charge potential, radii, 
and speeds. Then to calculate the averages seen in Table~\ref{plasmasing1},  
I sliced ($\pm$~0.1~\rj) through the radial locations:~6.2, 8, 10, 15, 20, and 30~\rj.
for the particles' sizes, charges and speeds.

In Table~\ref{plasmasing2}\marginpar{Table~\ref{plasmasing2}}, 
I list gyrofrequencies
calculated from Eqn.~(5), which includes a constant
magnetic field, and gyroradii calculated from Eqn.~(4) for ion, electron,
and dust particles.

\section{Towards Collective Particle Motions}    
 \label{towardscollective}

Collective interactions occur when self-generated fields of the
particles take over a correlative role in scattering the
electrons~\cite[Pg. 72]{Baum:1997}. The Debye screening length provides our first
indication of whether collective processes might be important. Finding
Debye lengths comparable to the spatial scales doesn't firmly
establish a collective plasma effect since it involves interactions
between individual particles and not between large groups of
particles in the plasma or the plasma as a whole. However, under the
presence of collisions, the particles might already behave differently from
what would be expected when using a single particle scenario. The
vacuum Coulomb force is long range, but in a plasma, this force only
extends a Debye length from the source, as a consequence of the
Debye shielding cloud~\cite[Pg. 37]{Crav:97}. Although on small
scales, a plasma in thermal equilibrium can have significant
departures from charge neutrality ($n_e \ne n_i$), for long spatial
scales, an equilibrium plasma must maintain charge neutrality. (quasi-neutrality.)

The size of the Debye length (i.e. shielding cloud) increases as the
electron temperature increases because electrons with greater
kinetic energy are better able to overcome the Coulomb attraction
associated with the potential (in the vicinity of the test charge).
On the other hand, the value $\lambda_D$ is smaller for a denser
plasma because more electrons are available to populate the
shielding cloud. (And note that the density variation is greater
when the electron gas is cold than when the gas is hot).

The Debye length for electrons is defined as \cite[Pg.
37]{Crav:97} :

$$\lambda _D = \sqrt{{ \epsilon_{\circ} k_B T_e} \over { q_e^2n_e}} \quad  {(\rm SI)} \quad \eqno(6) $$

If I determine that collective effects in a plasma are possible,
then a plasma motion that I want to identify is Langmuir
oscillations. Langmuir oscillations lead to an oscillation
frequency in a fully ionized plasma called the electron plasma
frequency $\omega_e$~\cite[Pg. 3]{Baum:1997}. If the quasi-neutrality 
of the plasma is disturbed by some
external force, then, because the electrons are much more mobile than the
heavier ions, they are accelerated in an attempt to restore the charge
neutrality. Due to their inertia, they will move back and forth
around some equilibrium position, resulting in fast collective
oscillations around the more massive ions. Note that the electron
plasma frequency here, refers to a plasma oscillation, rather than to a
propagating wave. The definition of the ion plasma frequency and the
electron plasma frequency is~\cite[Pg. 94-97]{Crav:97}:

$${\omega _e=\sqrt {{n_eq_e^2} \over {\epsilon_{\circ} m_e}}
   \quad \omega _{i}=\sqrt {{n_iq_i^2} \over {\epsilon_{\circ} m_i}}}
  \quad  \omega_p =\sqrt {\omega ^2_e+\omega ^2_i} 
\quad {\omega _d=\sqrt {{n_dq_e^2} \over {\epsilon_{\circ} m_d}}}\quad \eqno(7)$$

However, the electron plasma frequency dominates the total plasma frequency
because $m_i>>m_e$, therefore: $\omega_p \approx \omega _e$. Table~\ref{plasmafreqs}\marginpar{Table~\ref{plasmafreqs}}
gives Debye lengths (Eqn.~6) and frequencies (Eqn.~7) for
electron/ion $\omega_p$ plasma oscillations, and the equivalent dust plasma 
frequency $\omega _d$. For the latter, I invoke a dust particle mass density 1500~kg-m$^{-3}$,
spherical particle of average radii $a$ seen in the middle panel of Fig.~1, 
a Jovian streams number density~$n_d$, valid for distance 6~\rj\ -- 2~\AU, ranging from
10$^{-3}$ -- 10$^{-8}$~m$^{-3}$~\cite[Table~10.1]{Kru:2003}. Since $n_d d^3 = 1$,  then
the intergrain distance $d$ of the dust streams particles,
$d = \left( {{1 \over {n_d}}} \right)^{1/3}$, ranges from
10~--~500~m.

Now I can check `{Dust-in-Plasma}': $a \ll \lambda_D < d$ and `{Dusty
Plasma}':  \mbox{$a \ll d < \lambda_D$}. From the Debye lengths in
Table~\ref{plasmafreqs}, I see that collective effects might be
relevant in the magnetospheric region starting from roughly
10--15~\rj, where the intergrain distance $d$ is about 10~m and 
the Debye length is 10~m.  In other words,
outward from 10-15~\rj, in the Jovian magnetosphere, it is possible
that I have a ``dusty plasma", rather than a ``dust-in-plasma",
therefore more complex frequencies and waves might be seen here.

For more refined checks, I
examine selected Galileo orbits
which show higher quantities of dust present, in order to check dusty plasma conditions. 
Table~\ref{checkcriteria1}\marginpar{Table~\ref{checkcriteria1}} shows seven Galileo 
orbits: E4, G7, G8, E11, E15, C21, G28, which were chosen
from the 1996-2003 mission because they demonstrated a high number of impacts
plus a high mean dust impact rate. The fluxes  which determined the intergrain distance ($d$)
were average fluxes in which the sensitivity and viewing geometry of the dust detector were accounted, 
provided \mbox{by~\cite{Kru:2005}}. The speeds for each location at each Galileo encounter were average 
speeds calculated via the simulation described in the previous section, and seen in
the bottom panel of  Fig.~1.
The {\Large\bf $\rm *$} denotes regions where $d < \lambda_D$, i.e., of possible dusty plasma conditions.

\section{Havnes Parameter \protect{$P$} of a Dust Particle Embedded in Jupiter's Plasma}

The dimensionless parameter $P$ emerged from the simultaneous solution for the surface potential of an ensemble of grains
and the potential of the Maxwelli\-an energy-distributed plasma in which the grains were immersed; 
given ion and collection currents to the grains from the plasma, as well as photoelectron currents,
and assuming that the sum of the charging currents to the dust grain was zero. The solutions for the relative dust
potential $(U-V_p) {q_e / ({k_B T})}$ and the plasma potential $(V_p q_e) / ({k_B T})$ 
only depend on the parameter \citep{Havnes:87}:

$$P\equiv {{4\pi \epsilon_{\circ} a} \over {q_e^2n_e}}n_dk_BT \quad \eqno(8)$$

\noindent where $\epsilon _{\circ}$~=~the permittivity of free space (8.854~$\times$~10$^{-12}$~C$^2$-N$^{-1}$-m$^{-2}$), 
$a$~=~the dust grain radius~(m), $n_{d}$~=~dust number
density~(m$^{-3}$), $k_B$~=~the Boltzman constant (1.381~$\times$~10$^{-23}$~J-K$^{-1}$),
$T$ is the plasma temperature~(K) (Table~\ref{plasmareps}), $q_e$ is the charge on the electron (1.602~$\times$~10$^{-19}$~C),
and $n_e$ is the plasma density~(m$^{-3}$) (Table~\ref{plasmareps}).

The \citet[Fig.~1]{Havnes:90} solutions represent a general scenario of a distribution of dust particle sizes while
in plasmas of different composition. The scenario assumes equilibrium dust potentials and does not include secondary electron
emission currents. Earlier works by Havnes and colleagues considered more idealized assumptions for dust cloud geometry, dust particle
sizes, and plasma energy distributions, and note that
the definition for $P$ in the earlier works such as \citet{Goertz:89} must multiply $P$ by $4 \pi \epsilon_{\circ} / q_e$ 
in order to match Eqn.~(8).

The solutions from Havnes' work do not fit perfectly the charging and dynamics
of the Jovian dust streams for several reasons: 1) the secondary electron emission is an important 
charging  process for the dust stream particles, especially
in the inner Jovian magnetosphere \citep{Hor:97}, and 2) the dust stream particles do not have a unique 
equilibrium potential;   \citet{Mey:82}\  demonstrated that the equilibrium potential from 
secondary electron emission currents may have multiple roots. 
Moreover, an equilibrium potential is often not reached for the
smallest Jovian dust stream particles, where, for example, a 10~nm
sized particle needs
1--5 hours to reach equilibrium potential within the Jovian magnetosphere,
which is a significant part of the travel time for an escaping particle 
\cite{Graps:2001}.

Nevertheless, the potential solutions and parameter should be reliable `enough' indicators
of dusty plasma conditions. In Table~\ref{plasmahavnes}\marginpar{Table~\ref{plasmahavnes}},
I calculate $P$ for the same specific locations in Jupiter's magnetosphere
as Table~\ref{checkcriteria1}, considering temperature $T$~=~$T_e$,
given a dust number density $n_d= (1/d)^{(1/3)}$ from the intergrain distances $d$ in Table~\ref{checkcriteria1}, 
and applying the average sizes for each location at each Galileo encounter
that were calculated via the simulation described previously, and seen in
the middle panel of  Fig.~1.

When the Havnes parameter $P$ is very small, then the grain potential $U$
approaches the Spitzer single-grain value of -2.51~${k_BT_e}/q_e$, and
the plasma potential $V_p$ is zero. Then the grain can be treated as a test particle
and the grain's charge has a negligible effect on the plasma environment, and
on the electric and magnetic fields \citep{Hor:2004}.

The tiny $P$ values in Table~\ref{plasmahavnes}\ indicate no dusty plasma conditions, given the
present inputs to the Eqn.~(8).  The discrepancy between the two methods 
requires a discussion.

\section{Discussion}

The minimum requirement for collective effects is the
same requirement and expression for the existence of a plasma, where many particles are contained within 
a Debye sphere (to apply statistics), and for each
volume element, the plasma is electrically neutral,  but now we are substituting 
dust particles for (ions, electrons): ${({4\pi}/{3})} n_d \lambda^3_D$
(note added in proof). The minimum requirement can be expressed via a ratio of $d/\lambda_D =\kappa$, in
 \citet{Barkan:94}'s notation, where  $\kappa$ should be $< 1$. Several locations in
some of the Galileo spacecraft orbits satisfy the minimum requirement of a dusty plasma, but 
the conditions do not satisfy the necessary conditions for collective behavior.

A comparison of the energies of the dust streams and the plasma in which they are
embedded point to the main difficulty of the Jovian dust streams
displaying collective behavior (note added in proof). The plasma kinetic
energies are in the tens to approximately one hundred~eV, while the high energies of the dust streams 
are in the range of a few to hundreds of MeV.

The parameter $P$ is a more sensitive measure of collective behavior by giving the ratio 
for the charges on the dust compared to the electrons in the plasma. The potential solutions
as a function of $P$ \citep[Fig.~1]{Havnes:90} illustrate that 
collective effects of the dust ensemble occur when the parameter $P\sim$~1, which is when the derivative 
with respect to $P$ for the relative plasma potential caused by the dust, is at a maximum (note added in proof).
At this $P$, local electric potential differences of order $(k_BT/q_e)$~V can
be generated by dust density irregularities. If $P$ is less than one, then the maximum 
potential difference in the dusty plasma is smaller too \citep{Goertz:89,Havnes:90}.  
Therefore, if we have maximum collective effects, $P\sim$~1, then the maximum electrostatic 
potential difference  from the plasma given a typical temperature listed in Table~\ref{plasmareps}, 5~eV, is: 
$\left[ {U-V_p} \right]\left( {k_BT/q_e} \right)$~=~5~V. Since our
calculated $P$ is about 10$^{10}$ smaller than that, and likely 
smaller within any possible dust density irregularities, the true potential differences
will be likely very much smaller than 5~V.

It is useful to vary the dust and plasma number density contained in $P$ and see
the physical effects. If the conditions for dust favor collective behavior, then if 
the dust number density is decreased, the 
grain charges decrease. Here two effects play a role in opposite directions 
\cite[Pg. 28]{Ver:2000}:\ 1)~the capacitance of the grains increases, and 
2)~the mean charge for each grain decreases compared to the equilibrium charge 
of a single grain.  Therefore, an increase in dust density means that the
grain ensemble has a larger appetite for electrons, but that the number of 
available electrons per grain decreases. 

If the dust number density remains fixed, then from $n_d  Q\sim n_e q_e$ for maximum collective conditions, if the
dust charge density $n_dQ$ decreases, then the plasma number density $n_e$ 
should decrease (and therefore, $\lambda_D$ increases), as well. I follow with a quantitative example.

The ratio $\kappa$ (which gives the minimum requirement for a dusty plasma) can trace 
collective effects if properly compared to the number of charges on the grain $Z_d=Q/q_e$. 
I calculated $Z_d$ via the indirect measurement of $Q=\langle U \rangle 4 \pi \epsilon_{\circ} \langle a\rangle$, 
where $\langle U\rangle$ and $\langle a\rangle$ are the 
average potentials and sizes (output from the dust stream model, Fig.~1, as described previously). The ratio $\kappa$, which
compares the measured interparticle distance $d$ to the plasma Debye length $\lambda _D$, was 
calculated for each of the seven Galileo orbits at the same six locations, and I show them
both in Table~\ref{checkZd}\marginpar{Table~\ref{checkZd}}. In addition, I mark with an asterisk the
same locations where the {\it minimum} requirement for dusty plasma conditions ($\kappa < 1$) is satisfied.

The values in the table do not correlate the decrease of plasma density with dust charge. 
Two good locations in which to compare is between 6.2~\rj\ and 15~\rj, where the plasma density $n_e$ decreases drastically.
In orbits E4, G8, and G28, the dust number density $n_d$ is roughly constant at 
6.2~\rj\ and 15~\rj, and the ratios for $\kappa$ decreases. Yet $Z_d$ increases, not decreases, as it
should if there existed collective behavior. Moreover, a charge increase appeared
in a region: 20~\rj\ in orbit G8, where a comparison of Debye length and interparticle distance
showed possible dusty plasma conditions. Therefore, while it is useful to know the locations
in space and time for the minimum ($\kappa$) requirements of possible dusty plasma conditions, 
$P$ can still be small. To have collective behavior the conditions must be additionally supported by
the parameter $P$ closer to unity.

\section{Conclusions and Future Directions}

A natural laboratory to study electromagnetically-controlled cosmic dust has been
provided  by the Jovian dust streams and the data from the instruments which
were on board the Galileo spacecraft. Given the prodigious
quantity of dust poured into the Jovian magnetosphere by Io and its volcanoes 
resulting in the dust streams, the possibility of dusty plasma conditions exist.
To study the possibility, I showed how to distinguish between dust-in-plasma and
dusty-plasma and how the Havnes parameter $P$ can be used to support or negate 
the possibility of collective behavior of the dusty plasma. 

The result of applying these tools to the Jovian dust streams reveals mostly dust-in-plasma
behavior. In the orbits displaying the highest dust stream fluxes, portions
of orbits E4, G7, G8, C21 satisfy the minimum requirements for a dusty plasma. 
However, the $P$ parameter demonstrates that these mild dusty
plasma conditions do not lead to collective behavior of the dust stream
particles. This result might be a relief to the modelers who 
have treated the Jovian dust stream particles
as isolated particles in their charging and dynamics calculations
during the last decade. Or this result might motivate dust researchers
who yearn for collective dusty plasma conditions to look elsewhere 
in the solar system for their natural laboratories.

These other natural dusty plasma laboratories are conveniently
under study now or soon to be studied by other spacecraft.
One natural laboratory, the Saturn ring system (with its spokes),
has been the prime motivator for dusty plasma studies
since the time of the Voyager spacecraft twenty years ago. The 
Cassini spacecraft in orbit around Saturn since 2004 is well-placed
to continue those dusty plasma studies. Not only are the spokes 
today still the topic of PhD theses, but the active moon 
Enceladus is providing new dust puzzles.
In addition, comet missions such as Stardust in its flyby of comet 
81P/Wild 2,
and Rosetta, on its way to comet 67P/Churyumov-Gerasimenko, provide
the instruments and the dusty plasma source (comets) in which to study 
the complex interplay between dust and plasma.
 
Lest we not forget about Io, the most active volcanic dust source in the 
solar system, the archived Galileo data can
give information for Jupiter's magnetic field activity at the same 
time it can trace the activity of Io's volcanoes. This information 
might give new views for how the mass-loading
from Io influences the Jupiter environment. In future work, I 
would like to incorporate  real-time plasma and magnetic field data
with higher resolution DDS dust fluxes in order to continue to 
understand the Jovian dust streams and the Jupiter plasma and magnetosphere.



 

\vspace{.5cm}        
{\bf Software}

A spreadsheet that follows the calculations of this paper can be found
at:\\ {\footnotesize http://www.mpi-hd.mpg.de/dustgroup/$^{\mathrm \sim}$graps/dustyplasma/} .

\vspace{1cm}
{\bf Acknowledgments}

I thank Eberhard Gr\"un and Mihaly
Hor\'anyi for valuable discussions, Harald Kr\"uger
for good discussions, feedback and excellent Galileo DDS data support, 
two anonymous referees for their inestimable and important 
comments, the Galileo project at JPL
for effective and successful mission operations during
the Galileo spacecraft's years in orbit, and the INAF 
Istituto di Fisica dello Spazio Interplanetario for its
financial support.

\bibliographystyle{elsart-harv}
\bibliography{amthesis}

\begin{thebibliography}{23}
\expandafter\ifx\csname natexlab\endcsname\relax\def\natexlab#1{#1}\fi
\expandafter\ifx\csname url\endcsname\relax
  \def\url#1{\texttt{#1}}\fi
\expandafter\ifx\csname urlprefix\endcsname\relax\def\urlprefix{URL }\fi

\bibitem[{Bagenal(1989)}]{Bag:89}
Bagenal, F., 1989. Torus-magnetosphere coupling. In: Belton, M., West, R.~A.,
  Rahe, J. (Eds.), Time-Variable phenomena in the {J}ovian system. SP-494. pp.
  1--403.

\bibitem[{Barkan et~al.(1994)Barkan, {D'A}ngelo, , and Merlino}]{Barkan:94}
Barkan, A., {D'A}ngelo, N., , Merlino, R.~L., 1994. Charging of dust grains in
  a plasma. Physical Review Letters 73~(23), 3093--3096.

\bibitem[{Baumjohann and Treumann(1997)}]{Baum:1997}
Baumjohann, W., Treumann, R., 1997. Basic Space Plasma Physics. Imperial
  College Press.

\bibitem[{Cravens(1997)}]{Crav:97}
Cravens, T.~E., 1997. Physics of Solar System Plasmas. Cambridge University
  Press.

\bibitem[{Goertz(1989)}]{Goertz:89}
Goertz, C., 1989. Dusty plasmas in the solar system. Rev Geophys. 27, 271--292.

\bibitem[{Graps(1991)}]{Graps:1991}
Graps, A.~L., August 1991. Investigating the motions and energies of ions
  confined in a uniform magnetic field. Master's thesis, San Jose State
  University.
\newline\urlprefix\url{http://www.amara.com/ftpstuff/MSThesis.pdf}

\bibitem[{Graps(2001)}]{Graps:2001}
Graps, A.~L., Jul. 2001. {Io} revealed in the {J}ovian dust streams. Ph.D.
  thesis, Ruprecht-Karls-Universit{\"a}t Heidelberg.
\newline\urlprefix\url{http://www.ub.uni-heidelberg.de/archiv/1838}

\bibitem[{Graps et~al.(2000)Graps, Gr{\"u}n, Svedhem, Kr{\"u}ger, Hor{\'a}nyi,
  Heck, and Lammers}]{Graps:2000a}
Graps, A.~L., Gr{\"u}n, E., Svedhem, H., Kr{\"u}ger, H., Hor{\'a}nyi, M., Heck,
  A., Lammers, S., 2000. {I}o as a source of the {J}ovian dust streams. Nature
  405, 48--50.

\bibitem[{Gr{\"u}n et~al.(1996{\natexlab{a}})Gr{\"u}n, Baguhl, Hamilton,
  Riemann, Zook, Dermott, Fechtig, Gustafson, Hanner, Hor{\'a}nyi, Khurana,
  Kissel, Kivelson, Lindblad, Linkert, Linkert, Mann, McDonnell, Morfill,
  Polanskey, Schwehm, and Srama}]{Gru:96b}
Gr{\"u}n, E., Baguhl, M., Hamilton, D.~P., Riemann, R., Zook, H.~A., Dermott,
  S., Fechtig, H., Gustafson, B.~A., Hanner, M.~S., Hor{\'a}nyi, M., Khurana,
  K.~K., Kissel, J., Kivelson, M., Lindblad, B.-A., Linkert, D., Linkert, G.,
  Mann, I., McDonnell, J. A.~M., Morfill, G.~E., Polanskey, C., Schwehm, G.,
  Srama, R., May 1996{\natexlab{a}}. Constraints from {G}alileo observations on
  the origin of the {J}ovian dust streams. Nature 381, 395--398.

\bibitem[{Gr{\"u}n et~al.(1996{\natexlab{b}})Gr{\"u}n, Hamilton, Riemann,
  Dermott, Fechtig, Gustafson, Hanner, Heck, Hor{\'a}nyi, Kissel, Kr{\"u}ger,
  Lindblad, Linkert, Linkert, Mann, McDonnell, Morfill, Polanskey, Schwehm,
  Srama, and Zook}]{Gru:96a}
Gr{\"u}n, E., Hamilton, D.~P., Riemann, R., Dermott, S., Fechtig, H.,
  Gustafson, B.~A., Hanner, M.~S., Heck, A., Hor{\'a}nyi, M., Kissel, J.,
  Kr{\"u}ger, H., Lindblad, B.-A., Linkert, D., Linkert, G., Mann, I.,
  McDonnell, J. A.~M., Morfill, G.~E., Polanskey, C., Schwehm, G., Srama, R.,
  Zook, H.~A., 1996{\natexlab{b}}. Dust measurements during the initial
  {G}alileo {J}upiter approach and {I}o encounter. Science 274, 399--401.

\bibitem[{Gr{\"u}n et~al.(1997)Gr{\"u}n, Kr{\"u}ger, Graps, Hamilton, Heck,
  Dermott, Fechtig, Zook, Gustafson, Hanner, Hor{\'a}nyi, Kissel, Lindblad,
  Linkert, Linkert, Mann, McDonnell, Morfill, Polanskey, Schwehm, and
  Srama}]{Gru:97}
Gr{\"u}n, E., Kr{\"u}ger, H., Graps, A.~L., Hamilton, D.~P., Heck, A., Dermott,
  S., Fechtig, H., Zook, H.~A., Gustafson, B.~A., Hanner, M.~S., Hor{\'a}nyi,
  M., Kissel, J., Lindblad, B.~A., Linkert, D., Linkert, G., Mann, I.,
  McDonnell, J. A.~M., Morfill, G.~E., Polanskey, C., Schwehm, G., Srama, R.,
  Sep. 1997. Dust measurements in the {J}ovian magnetosphere. Geophys. Res.
  Letters 24, 2171--2174.

\bibitem[{Havnes et~al.(1990)Havnes, Aanesen, and Melands{\o}}]{Havnes:90}
Havnes, O., Aanesen, T.~K., Melands{\o}, F., 1990. On dust charges and plasma
  potentials in a dusty plasma with dust size distribution. J. Geophys. Res.
  95~(A5), 6581--8585.

\bibitem[{Havnes et~al.(1987)Havnes, Goertz, Morfill, Gr{\"u}n, and
  Ip}]{Havnes:87}
Havnes, O., Goertz, C., Morfill, G., Gr{\"u}n, E., Ip, W., 1987. Dust charges,
  cloud potential and instabilities in a dust cloud embedded in a plasma. J.
  Geophys. Res. 92~(A3), 2281--2287.

\bibitem[{Hor{\'a}nyi and Cravens(1996)}]{Hor:96b}
Hor{\'a}nyi, M., Cravens, T., 1996. The structure and dynamics of {J}upiter's
  ring. Nature 381, 293--295.

\bibitem[{Hor{\'a}nyi et~al.(1997)Hor{\'a}nyi, Gr{\"u}n, and Heck}]{Hor:97}
Hor{\'a}nyi, M., Gr{\"u}n, E., Heck, A., 1997. Modeling the {G}alileo dust
  measurements at {J}upiter. Geophys. Res. Letters 24, 2175--2178.

\bibitem[{{Hor{\'a}nyi} et~al.(2004){Hor{\'a}nyi}, {Hartquist}, {Havnes},
  {Mendis}, and {Morfill}}]{Hor:2004}
{Hor{\'a}nyi}, M., {Hartquist}, T.~W., {Havnes}, O., {Mendis}, D.~A.,
  {Morfill}, G.~E., Dec 2004. Dusty plasma effects in {S}aturn's magnetosphere.
  Reviews of Geophysics 42, RG4002.

\bibitem[{{Krivov} et~al.(2002){Krivov}, {Kr{\"u}ger}, {Gr{\"u}n},
  {Thiessenhusen}, and {Hamilton}}]{Kriv:2000}
{Krivov}, A.~V., {Kr{\"u}ger}, H., {Gr{\"u}n}, E., {Thiessenhusen}, K.,
  {Hamilton}, D.~P., Jan. 2002. {A tenuous dust ring of {J}upiter formed by
  escaping ejecta from the {G}alilean satellites}. Journal of Geophysical
  Research (Planets) 107, 1029--1041.

\bibitem[{Kr{\"u}ger(2005)}]{Kru:2005}
Kr{\"u}ger, H., 2005. personal communication.

\bibitem[{{Kr{\"u}ger} et~al.(2003){Kr{\"u}ger}, {Hor{\'a}nyi}, and
  {Gr{\"u}n}}]{Kru:2003c}
{Kr{\"u}ger}, H., {Hor{\'a}nyi}, M., {Gr{\"u}n}, E., Jan. 2003. {{J}ovian dust
  streams: Probes of the {I}o plasma torus}. Geophys. Res. Letters 30,
  1058--1061.

\bibitem[{Kr{\"u}ger et~al.(2004)Kr{\"u}ger, Hor\'anyi, Krivov, and
  Graps}]{Kru:2003}
Kr{\"u}ger, H., Hor\'anyi, M., Krivov, A., Graps, A., 2004. Jovian dust:
  Streams, clouds and rings. In: Jupiter: The Planet, Satellites \&
  Magnetosphere. Cambridge University Press, pp. 219--240.

\bibitem[{{Meyer-Vernet}(1982)}]{Mey:82}
{Meyer-Vernet}, M., 1982. {`Flip-flop'} of electric potential of dust grains in
  space. Astr. Ap. 105, 98--106.

\bibitem[{Spencer and Schneider(1996)}]{Spens:96}
Spencer, J.~R., Schneider, N.~M., 1996. {I}o on the eve of the {G}alileo
  mission. Ann. Rev. Earth Sp. Sci. 24, 125--190.

\bibitem[{Verheest(2000)}]{Ver:2000}
Verheest, F., 2000. Waves in Dusty Space Plasmas. Kluwer.

\end{thebibliography}
%

\newpage
\SSM

\begin{table}[ht]
\caption [Jupiter Plasma Representative Numbers]{Jupiter Plasma Representative Numbers}
\begin{tabular}{|llllll|}\hline
Location  & $T_e$  & $T_i$   & $n_e$  & avim & $m_i$  \\
(\rj) & (eV) & (eV) & (m$^{-3}$) &  (a.m.u.) & (kg) \rule{0in}{3ex} \\[1ex]\hline\hline
6.2 & 4.9 & 60 & 1.3~$\times$~10$^{9}$  & 22.5 & 3.7~$\times$~10$^{-26}$ \rule{0in}{3ex} \\
8 & 5.4 & 60 & 1.6~$\times$~10$^{8}$ & 19.6 &  3.2~$\times$~10$^{-26}$ \\
10 & 22.5 & 100 & 2.3~$\times$~10$^{7}$   & 17.1  & 2.8~$\times$~10$^{-26}$ \\
15 & 22.0 & 120 & 3.6~$\times$~10$^{6}$  & 15.6 & 3.7~$\times$~10$^{-26}$  \\
20 & 22.0 & 120 & 1.3~$\times$~10$^{6}$   & 15.6  & 2.6~$\times$~10$^{-26}$ \\
30 & 22.0 & 120 & 1.0~$\times$~10$^{6}$   & 15.6  & 2.6~$\times$~10$^{-26}$ \\ 
\hline
\multicolumn{6}{l}{{\footnotesize \rj: distance from the center of Jupiter = 71398 km}}\\ 
\multicolumn{6}{l}{{\footnotesize  $T_e$: electron temperature, $T_i$: ion temperature}}\\
\multicolumn{6}{l}{{\footnotesize   $n_e$: electron number density, avim: average ion mass, $m_i$: ion mass}}\\
\end{tabular}
\label{plasmareps}
\end{table}

\vspace{.3cm}

\begin{table}[htpb]
\caption [Jupiter Plasma Single Particle Motions I]{Jupiter Plasma Single Particle Motions Part I}
\begin{tabular}{|llllll|}\hline
Location  & $a$  & $|Q_d|  $ & $\langle v_i\rangle$  & $\langle v_e\rangle$   & $\langle v_d\rangle$  \\
{ (\rj)}& (m) & (C)  & (m-s$^{-1}$)  & (m-s$^{-1}$)   & (m-s$^{-1}$)  \rule{0in}{1ex}\\[1ex]\hline\hline 
6.2 &1.2~$\times$~10$^{-8}$ & 7.3~$\times$~10$^{-18}$ &   2.3~$\times$~10$^4$ & 1.3~$\times$~10$^6$ & 5.2~$\times$~10$^4$ \rule{0in}{3ex} \\
8 & 1.2~$\times$~10$^{-8}$ & 3.3~$\times$~10$^{-18}$ &  2.4~$\times$~10$^4$ & 1.4~$\times$~10$^6$ & 1.2~$\times$~10$^5$ \\
10 & 1.8~$\times$~10$^{-8}$ & 9.1~$\times$~10$^{-18}$ &  3.4~$\times$~10$^4$ & 2.8~$\times$~10$^6$ & 9.9~$\times$~10$^4$ \\
15 &  2.4~$\times$~10$^{-8}$ & 1.5~$\times$~10$^{-17}$ & 3.8~$\times$~10$^4$ & 2.8~$\times$~10$^6$ & 8.2~$\times$~10$^4$  \\
20 & 2.4~$\times$~10$^{-8}$ & 1.5~$\times$~10$^{-17}$ & 3.8~$\times$~10$^4$ & 2.8~$\times$~10$^6$ & 9.1~$\times$~10$^4$ \\
30 & 2.7~$\times$~10$^{-8}$ &2.0~$\times$~10$^{-17}$ & 3.8~$\times$~10$^4$ & 2.8~$\times$~10$^6$ & 1.1~$\times$~10$^5$ \\ \hline
\end{tabular}
\label{plasmasing1}
\end{table}

\vspace{.3cm}

\begin{table}[htbp]
\caption [Jupiter Plasma Single Particle Motions II]{Jupiter Plasma Single Particle Motions Part II}
\begin{center}
\begin{tabular}{|lllllll|}\hline
Location  & $\Omega_i$  & $\Omega_e$   & $\Omega_d$  & $R_{gi}$   & $R_{ge}$ & $R_{gd}$  \\
(\rj) & (sec$^{-1}$) & (sec$^{-1}$)  & (sec$^{-1}$) & (m) & (m) & (m) \rule{0in}{3ex} \\[1ex] \hline\hline
6.2 & 7.5 & 3.1~$\times$~10$^5$ & 1.2~$\times$~10$^{-3}$ & 3.0~$\times$~10$^3$ & 4.2  & 4.4~$\times$~10$^7$\rule{0in}{3ex} \\
8 & 4.0 & 1.4~$\times$~10$^5$ & 2.5~$\times$~10$^{-4}$ & 6.0~$\times$~10$^3$ & 9.6  & 4.7~$\times$~10$^8$\\
10 & 2.4  & 7.4~$\times$~10$^4$ & 1.0~$\times$~10$^{-4}$ & 1.4~$\times$~10$^4$ & 3.8~$\times$~10$^1$ & 9.6~$\times$~10$^8$\\
15 & 7.7~$\times$~10$^{-1}$ & 2.2~$\times$~10$^4$ & 2.1~$\times$~10$^{-5}$ & 5.0~$\times$~10$^4$ &  1.3~$\times$~10$^2$  & 3.9~$\times$~10$^9$ \\
20 & 3.2~$\times$~10$^{-1}$ & 9.2~$\times$~10$^3$ & 9.3~$\times$~10$^{-6}$ & 1.2~$\times$~10$^5$ &  3.0~$\times$~10$^2$ & 9.8~$\times$~10$^9$\\
30 & 9.6~$\times$~10$^{-2}$ & 2.7~$\times$~10$^3$ & 2.5~$\times$~10$^{-6}$ & 4.0~$\times$~10$^5$ &  1.0~$\times$~10$^3$ & 4.6~$\times$~10$^{10}$\\ \hline
\end{tabular}
\end{center}
\label{plasmasing2}
\end{table}

\newpage

\begin{center}
\begin{table}[htbp]
\caption [Jupiter Plasma Debye Lengths and Frequencies]{Jupiter Plasma Debye Lengths and Frequencies}
\begin{tabular}{|lllc|}\hline
Location  & $\lambda_D$  & $\omega_p$   & $\omega_d$  \\
(\rj) & (m) & (s$^{-1}$) & (s$^{-1}$) \rule{0in}{3ex} \\[1ex]\hline\hline
6.2 & 4.5~$\times$~10$^{-1}$ & 2.0~$\times$~10$^6$ & 7.4~$\times$~10$^{-4}$ -- 7.4~$\times$~10$^{-7}$  \rule{0in}{3ex} \\
8 & 1.4  & 7.0~$\times$~10$^5$ & 3.3~$\times$~10$^{-4}$ -- 3.3~$\times$~10$^{-7}$ \\
10 & 7.3  & 2.7~$\times$~10$^5$ &   5.0~$\times$~10$^{-4}$ -- 5.0~$\times$~10$^{-7}$  \\
15 & 1.8~$\times$~10$^{1}$ & 1.1~$\times$~10$^5$ & 5.2~$\times$~10$^{-4}$ -- 5.2~$\times$~10$^{-7}$   \\
20 & 3.0~$\times$~10$^{1}$ & 6.4~$\times$~10$^4$ & 5.5~$\times$~10$^{-4}$ -- 5.5~$\times$~10$^{-7}$  \\
30 & 3.5~$\times$~10$^{1}$ & 5.6~$\times$~10$^4$ & 6.0~$\times$~10$^{-4}$ -- 6.0~$\times$~10$^{-7}$  \\ \hline
\end{tabular}
\label{plasmafreqs}
\end{table}
\end{center}

\vspace{.3cm}

\begin{center}
\begin{table}[htpb]
\caption [Selected Galileo Orbits]{Checking Interparticle Distance \protect{$d$} for Selected Galileo Orbits}
\begin{tabular}{|llllllll|}\hline
{ r}  &  {\rm E4} & {\rm G7} &  {\rm G8} & {\rm E11 } & {\rm E15  }&  {\rm C21} & {\rm G28 } \\ 
{ (\rj)} &  { $d$ (m)} & { $d$ (m)} &  {$d$ (m)}  & {$d$ (m)} & {$d$ (m) }&  {$d$ (m)} & {$d$ (m)} \rule{0in}{1ex} \\[1ex]\hline\hline
6.2  &1.2~$\times$~10$^{2}$ &  8.0~$\times$~10$^{1}$ & 5.6~$\times$~10$^{1}$ & 1.6~$\times$~10$^{2}$ &  1.2~$\times$~10$^{2}$ &  4.7~$\times$~10$^{1}$ & 1.2~$\times$~10$^{2}$   \\
8  & 5.8~$\times$~10$^{3}$ &   1.0~$\times$~10$^{2}$ &  7.3~$\times$~10$^{1}$  &  2.1~$\times$~10$^{2}$ & 1.6~$\times$~10$^{2}$ & 5.8~$\times$~10$^{1}$ &	1.6~$\times$~10$^{2}$ \\
10 & 4.6~$\times$~10$^{3}$ & 7.9~$\times$~10$^{1}$ & 7.9~$\times$~10$^{1}$ &  1.7~$\times$~10$^{2}$ & 1.3~$\times$~10$^{2}$ &  2.7~$\times$~10$^{1}$ & 1.5~$\times$~10$^{2}$ \\
15   & 3.4~$\times$~10$^{2}$  & 2.2~$\times$~10$^{1}$ &  3.4~$\times$~10$^{1}$ & 7.4~$\times$~10$^{1}$ & 2.9~$\times$~10$^{1}$ &  8.0 *& 1.6~$\times$~10$^{2}$\\
20  &1.8~$\times$~10$^{2}$  &   2.1~$\times$~10$^{1}$ * & 2.6~$\times$~10$^{1}$ *   &  4.5~$\times$~10$^{1}$ & 4.5~$\times$~10$^{1}$ & 	7.1 *  &  1.7~$\times$~10$^{2}$ \\
30 & 3.1~$\times$~10$^{1}$ * &   8.3~$\times$~10$^{1}$ & 3.1~$\times$~10$^{1}$ * &  7.2~$\times$~10$^{1}$ &  9.6~$\times$~10$^{1}$ & 2.8~$\times$~10$^{1}$ *  &  2.2~$\times$~10$^{2}$\\
\hline
\end{tabular}  
\label{checkcriteria1}
\end{table}
\end{center}

\vspace{.3cm}

\begin{center}
\begin{table}[htpb]
\caption [Checking Havnes Parameter]{Checking Havnes Parameter \protect{$P$} for Selected Galileo Orbits}
\begin{tabular}{|llllllll|}\hline
{ $r$}  &{\rm E4} & {\rm G7} &  {\rm G8} & {\rm E11 } & {\rm E15 } &  {\rm C21} & {\rm G28 } \rule{0in}{1ex} \\
{ (\rj)} &  $P$  & $P$ & $P$   & $P$ & $P$ &  $P$ &  $P$ \rule{0in}{1ex} \\[1ex]\hline\hline
6.2 & 1.8$\times$10$^{-14}$	&6.0$\times$10$^{-14}$	&1.8$\times$10$^{-13}$	&7.2$\times$10$^{-15}$	&1.8$\times$10$^{-14}$	&3.0$\times$10$^{-13}$	&1.8$\times$10$^{-14}$\\
8 & 1.5$\times$10$^{-15}$	&2.5$\times$10$^{-13}$	&7.4$\times$10$^{-13}$	&3.0$\times$10$^{-14}$	&7.4$\times$10$^{-14}$	&1.5$\times$10$^{-12}$	&7.4$\times$10$^{-14}$\\
10 & 1.2$\times$10$^{-13}$	&2.4$\times$10$^{-11}$	&2.4$\times$10$^{-11}$ &2.4$\times$10$^{-12}$	&6.1$\times$10$^{-12}$	&6.1$\times$10$^{-10}$ & 3.7$\times$10$^{-12}$\\
15 & 2.5$\times$10$^{-9}$ 	&1.0$\times$10$^{-8}$	& 2.5$\times$10$^{-9}$ &	2.5$\times$10$^{-10}$	&4.4$\times$10$^{-9}$ &2.0$\times$10$^{-7}$ 	&2.5$\times$10$^{-11}$\\
20 & 4.9$\times$10$^{-8}$ 	&3.1$\times$10$^{-8}$ 	&1.5$\times$10$^{-8}$	&3.1$\times$10$^{-9}$ 	&3.1$\times$10$^{-9}$ 	&7.7$\times$10$^{-7}$ 	&6.2$\times$10$^{-11}$\\
30 & 1.4$\times$10$^{-8}$ &7.2$\times$10$^{-10}$	&1.4$\times$10$^{-8}$	&1.1$\times$10$^{-9}$	&4.7$\times$10$^{-10}$	&1.8$\times$10$^{-8}$	&3.6$\times$10$^{-11}$\\
\hline
\end{tabular}  
\label{plasmahavnes}
\end{table}
\end{center}

\newpage

\begin{table}[htpb]
\caption [\protect{$\kappa$} Change with \protect{$Z_d$}]{Checking If \protect{$\kappa$} Changes With \protect{$Z_d$} for Selected Galileo Orbits}
\begin{tabular}{|lllllllll|}\hline
{ r}  & $Z_d$  &  {\rm E4} & {\rm G7} &  {\rm G8} & {\rm E11 } & {\rm E15  }&  {\rm C21} & {\rm G28 } \\ 
{ (\rj)} & & $\kappa$  & $\kappa$ & $\kappa$   & $\kappa$ & $\kappa$ &  $\kappa$ &  $\kappa$ \rule{0in}{1ex} \\[1ex]\hline\hline
6.2  &45 &  54 & 36 & 25 &  73 & 54  & 21 & 54   \\
8  & 20 & 4200  &  77 &  53  &  160 & 110 & 42 &	110 \\
10 & 57& 640 & 11 & 11 &   24 & 17 &  3.8 & 21\\
15   &91  &   19 &1.2   &  1.9 & 4.0 & 	1.6 & 	0.43 *  &  8.7   \\
20  &96  &   5.9 &0.69 *&  0.87 * & 1.5 & 	1.5 & 	0.24 * &  5.5   \\
30 & 120 &   0.80 * & 2.2 &  0.80 * &  1.9 & 2.5  &  0.74 *  &  5.9\\
\hline
\end{tabular}  
\label{checkZd}
\end{table}

\vfil\eject

\newpage
\DDS
\thispagestyle{empty}	

\begin{figure}\label{ringrelease}
\epsfysize=0.95\vsize
\hspace{1in}
\epsfbox{RingChargeSizeSpeed.epsf}
\end{figure}

\vfil\eject
\newpage

{\bf Figure Caption}

Useful parameters from a simulation of of dust stream
particles  at 15 hours after release from a radius 6.2~\rj\ ring.
{\bf Top)}~Charge potential (in V), {\bf Middle)}~Particle size (in nm), and 
{\bf Bottom)}~Speed versus radius from the center of 
Jupiter in units of Jupiter radius (distance \rj = 71492~km).  
The (top) charge potential
panel illustrates the positive charging of the dust especially near the plasma torus 
due to the secondary electron 
emission current \citep{Hor:97}. 
 The middle panel illustrates the window of ejection sizes
\citep{Hor:96b,Kriv:2000,Graps:2001}, where particles smaller than a particular size
gyrate along Jupiter's magnetic field lines and never escape, while particles
larger than a particular size are dominated by gravitational forces and unable
to escape. The dearth of particle sizes in the middle of the panel are those 
whose sizes were favorable to ejection from the magnetosphere. The speeds seen in
the bottom panel demonstrate that most of the fastest particles particles ($>$~400~\kms)
escaped quickly, a few larger particles ($>$~20nm), still escaping can be seen 
scattered in speed in the middle of the figure. The rest of the particles are
bound, one population, seen along a steep speed gradient from about 300~\kms\ down to
80~\kms, are those that are dominated by Lorentz forces, while the others, along the bottom of the figure
that show a gentle speed gradient, are dominated by gravitational forces.

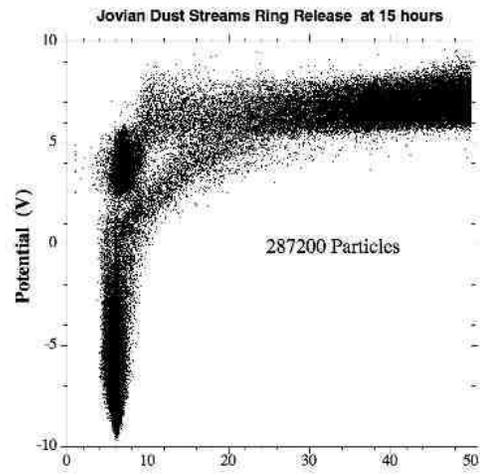
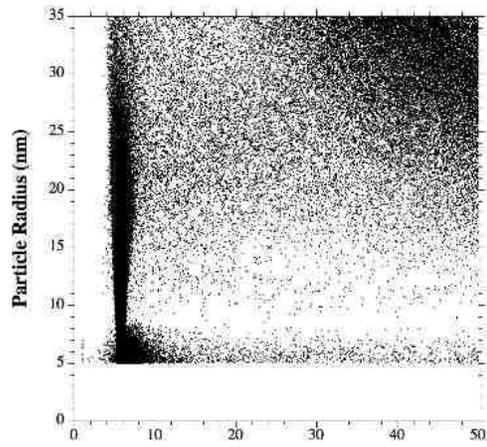
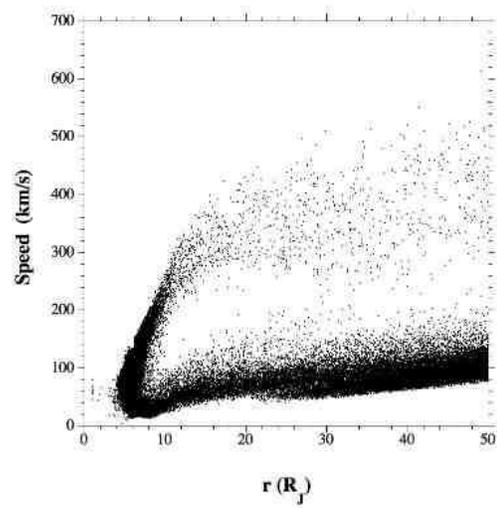
\end{document}